\documentclass[aps,prl,twocolumn,floatfix,nofootinbib,showpacs,superscriptaddress]{revtex4-1}

\usepackage{amsmath,amsfonts,amssymb,bm}
\usepackage{graphicx}
\usepackage{color}
\usepackage{pzccal}
\usepackage{soul}
\definecolor{purple}{rgb}{0.5,0,0.5}
\definecolor{blue}{rgb}{0.0,0,0.9}

\begin{document}

\title{Revealing dressed-quarks via the proton's charge distribution}

\author{Ian C.~Clo\"et}
\affiliation{CSSM and CoEPP, School of Chemistry and Physics
University of Adelaide, Adelaide SA 5005, Australia}
\affiliation{Physics Division, Argonne National Laboratory, Argonne, Illinois 60439, USA}

\author{Craig D.~Roberts}
\affiliation{Physics Division, Argonne National Laboratory, Argonne, Illinois 60439, USA}
\affiliation{Department of Physics, Illinois Institute of Technology, Chicago, Illinois 60616-3793, USA}

\author{Anthony W. Thomas}
\affiliation{CSSM and CoEPP, School of Chemistry and Physics
University of Adelaide, Adelaide SA 5005, Australia}

\date{3 April 2013}

\begin{abstract}
The proton is arguably the most fundamental of Nature's readily detectable building blocks.  It is at the heart of every nucleus and has never been observed to decay.  It is nevertheless a composite object, defined by its valence-quark content: $u+u+d$ -- i.e., two up ($u$) quarks and one down ($d$) quark; and the manner by which they influence, \emph{inter alia}, the distribution of charge and magnetisation within this bound-state.
Much of novelty has recently been learnt about these distributions; and it now appears possible that the proton's momentum-space charge distribution possesses a zero.  Experiments in the coming decade should answer critical questions posed by this and related advances; and we explain how such new information may assist in charting the origin and impact of key emergent phenomena within the strong interaction.  Specifically, we show that the possible existence and location of a zero in the proton's electric form factor are a measure of nonperturbative features of the quark-quark interaction in the Standard Model, with particular sensitivity to the running of the dressed-quark mass.
\end{abstract}

\pacs{
13.40.Gp, 	
14.20.Dh,	
12.38.Aw,   
12.38.Lg   
}

\maketitle

Experiments during the last decade have imposed a new ideal.  Namely, despite its simple valence-quark content, the proton's internal structure is very complex, with marked differences between the distributions of charge and magnetisation.  The challenge now is to explain the observations in terms of elemental nonperturbative features of the strong interaction.
In this connection, we demonstrate herein that the behaviour of the proton's electric form factor in the $6\,$-$10\,$GeV$^2$ range is particularly sensitive to the rate at which the dressed-quark mass runs from the nonperturbative into the perturbative domain of quantum chromodynamics (QCD), the strong interaction sector of the Standard Model.

The proton's momentum-space charge and magnetisation distributions are measured through combinations of the two Poincar\'e-invariant elastic form factors that are required to express the proton's electromagnetic current:
\begin{equation}
i e \, \bar u(p^\prime) \big[ \gamma_\mu F_1(Q^2) +
\frac{Q_\nu}{2m_N}\, \sigma_{\mu\nu}\,F_2(Q^2)\big] u(p)\,,
\end{equation}
where $Q=p^\prime - p$, $u(p)$ and $\bar u(p^\prime)$ are, respectively, spinors describing the incident, scattered proton, and $F_{1,2}(Q^2)$ are the proton's Dirac and Pauli form factors.  The charge and magnetisation distributions \cite{Sachs:1962zzc}
\begin{eqnarray}
\label{GEpeq}
G_E(Q^2)  &=&  F_1(Q^2) - \tau F_2(Q^2)\,,\;\\
\label{GMpeq}
G_M(Q^2)  &=&  F_1(Q^2) + F_2(Q^2)\,,
\end{eqnarray}
feature in the electron-proton scattering cross-section
\begin{equation}
\left(\frac{d\sigma}{d\Omega}\right) =
\left(\frac{d\sigma}{d\Omega}\right)_{\rm Mott}
\left[ G_E^2(Q^2) + \frac{\tau}{\varepsilon} G_M^2(Q^2)\right]
\frac{1}{1+\tau} \,,
\label{eq:rosenbluthM}
\end{equation}
where $\tau = Q^2/[4 m_N^2]$, $m_N$ is the proton's mass, and $\varepsilon$ is the polarisation of the virtual photon that mediates the interaction in Born approximation.  

The first data on the proton's form factors were made available by the experiments described in Ref.\,\cite{Hofstadter:1955ae}.  In Born approximation one may infer the individual contribution from each form factor to the cross section by using the technique of Rosenbluth separation \cite{Rosenbluth:1950yq}.  Namely, one considers the reduced cross-section, $\sigma_{\rm R}$, defined via:
\begin{equation}
\sigma_{\rm R} \, \left(\frac{d\sigma}{d\Omega}\right)_{\rm Mott}
:= \varepsilon (1+\tau) \, \frac{d\sigma}{d\Omega}\,.
\label{eq:sigmareduced}
\end{equation}
It is plain from Eq.\,\eqref{eq:rosenbluthM} that $\sigma_{\rm R}$ is linearly dependent on $\varepsilon$; and so a linear fit to the reduced cross-section, at fixed $Q^2$ but a range of $\varepsilon$ values, provides $G_E^2(Q^2)$ as the slope and $\tau G_M^2(Q^2)$ as the $\varepsilon=0$ intercept.  Owing to the relative factor of $\tau$, however, the signal for $G_M^2(Q^2)$ is enhanced with increasing momentum transfer, a fact which complicates an empirical determination of the proton's charge distribution for $Q^2\gtrsim 1\,$GeV$^2$.  Notwithstanding this, of necessity the method was employed exclusively until almost the turn of the recent millennium and, on a domain that extends to 6\,GeV$^2$, it produced
\begin{equation}
\label{GEeqGM}
\left. \mu_p\, \frac{G_E(Q^2)}{G_M(Q^2)} \right|_{\rm Rosenbluth} \approx 1\,,
\end{equation}
and hence a conclusion that the distributions of charge and magnetisation within the proton are approximately identical on this domain \cite{Arrington:2006zm,Perdrisat:2006hj}.  Significantly, this outcome is consistent with the, then popular, simple pictures of the proton's internal structure in which, e.g., quark orbital angular momentum and correlations play little role.

The situation changed dramatically when the combination of high energy, current and polarisation at the Thomas Jefferson National Accelerator Facility enabled polarisation-transfer reactions to be measured \cite{Jones:1999rz}.  In Born approximation, the scattering of longitudinally polarised electrons results in a transfer of polarisation to the recoil proton with only two nonzero components: P$_\perp$, perpendicular to the proton momentum in the scattering plane; and P$_\parallel$, parallel to that momentum.  The ratio ${\rm P}_\perp/{\rm P}_\parallel$ is proportional to $G_E(Q^2)/G_M(Q^2)$ \cite{Akhiezer:1974em,Arnold:1980zj}.  A series of such experiments
\cite{Jones:1999rz,Gayou:2001qd,Gayou:2001qt,Puckett:2010ac,Puckett:2011xg}
has determined that $G_E(Q^2)/G_M(Q^2)$ decreases almost linearly with $Q^2$ and might become negative for $Q^2 \gtrsim 8\,$GeV$^2$.  Such behaviour contrasts starkly with Eq.\,\eqref{GEeqGM}; and since the proton's magnetic form factor is reliably known on a spacelike domain that extends to $Q^2 \approx 30\,$GeV$^2$ \cite{Kelly:2004hm,Bradford:2006yz}, the evolution of this ratio exposes novel features of the proton's charge distribution, as expressed in $G_E(Q^2)$.

An explanation of the discrepancy between the Rosenbluth and polarisation transfer results for the ratio is currently judged to lie in two-photon-exchange corrections to the Born approximation, which affect the polarisation transfer extraction of $G_E(Q^2)/G_M(Q^2)$ far less than they do the ratio inferred via Rosenbluth separation \cite{Arrington:2011dn}.  The last decade has thus forced acceptance of a new paradigm; viz., the proton's internal structure must actually be very complex, with marked differences between the distributions of charge and magnetisation.

Given that sixty years of experimental effort has thus far discovered only one hadronic form factor that displays a zero; namely, the Pauli form factor associated with the transition between the proton and its first radial excitation (the Roper resonance), and that this feature was discovered just recently \cite{Dugger:2009pn,Aznauryan:2009mx,Aznauryan:2011td}, the chance that the proton's electric form factor might become negative is fascinating.  It is therefore worth elucidating the conditions under which that outcome is realisable \emph{before} the zero is empirically either located or eliminated as a reasonable possibility.  This is even more valuable if the appearance or absence of a zero is causally connected with a fundamental nonperturbative feature of the Standard Model.

Consider therefore a continuum computation of the proton's elastic form factors.  This has been accomplished within the Dyson-Schwinger equation (DSE) framework \cite{Bashir:2012fs}, which is distinguished by the feature that its elements have a direct connection with QCD. 

To illustrate this point, we note that QCD's dressed-quark propagator has the form
\begin{equation}
S(p) = 1/[i\gamma\cdot p A(p^2) + B(p^2)]\,,
\label{eqSp}
\end{equation}
where $Z(p^2)=1/A(p^2)$ is the wave-function renormalisation function and $M(p^2)=B(p^2)/A(p^2)$ is the renormalisation-point-invariant dressed-quark mass function.  In QCD with massless current-quarks, any finite-order perturbative computation yields $M(p^2)\equiv 0$.  However, a nonperturbative solution of the DSE for the dressed-quark propagator (QCD's gap equation) predicts a nonzero mass function with a strong momentum dependence \cite{Bhagwat:2006tu,Roberts:2007ji} -- a prediction confirmed by simulations of lattice-regularised QCD \cite{Bowman:2005vx}, so that it is now theoretically established  that chiral symmetry is dynamically broken in QCD.  The origin of the vast bulk of the mass of visible matter in the Universe therefore lies in the emergent strong-interaction phenomena of dynamical chiral symmetry breaking (DCSB) and confinement \cite{national2012Nuclear}. 

At the subnuclear level, DCSB has far-reaching consequences for meson properties \cite{Bashir:2012fs} and must be expected to impact just as heavily on baryons.  To expose novel aspects of this, we first recall that the proton is a bound-state in quantum field theory.  As such, its structure is described by a Faddeev amplitude, $\Psi$, obtained from a Poincar\'e-covariant Faddeev equation \cite{Cahill:1988dx}, which sums all possible quantum field theoretical exchanges and interactions that can take place between the three quarks that define its valence-quark content.  With $\Psi$ in hand, the proton's elastic form factors may be computed once the associated electromagnetic current is determined.

\begin{figure}[t]
\centerline{%
\includegraphics[clip,width=0.49\textwidth]{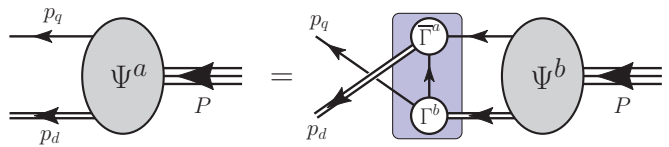}}
\caption{\label{figFaddeev} Poincar\'e covariant Faddeev equation.  $\Psi$ is the Faddeev amplitude for a proton of total momentum $P= p_q + p_d$.  The shaded rectangle demarcates the kernel of the Faddeev equation: \emph{single line}, dressed-quark propagator; $\Gamma$,  diquark correlation amplitude; and \emph{double line}, diquark propagator.}
\end{figure}

A dynamical prediction of Faddeev equation solutions obtained with the realistic interactions that describe the dressed-quark mass function, is the appearance of nonpointlike quark$+$quark (diquark) correlations within the proton \cite{Cahill:1987qr,Maris:2002yu}.  Whether one exploits this feature in developing an approximation to the quark-quark scattering matrix within the Faddeev equation \cite{Cloet:2008re,Chang:2011tx,Cloet:2011qu}, as illustrated in Fig.\,\ref{figFaddeev}, or chooses instead to eschew the simplification it offers, the outcome, when known, is the same \cite{Eichmann:2009qa}.  Notably, empirical evidence in support of the presence of diquarks in the proton is accumulating \cite{Close:1988br,Cloet:2005pp,Wilson:2011aa,Cates:2011pz,Cloet:2012cy}.

For a proton described by the amplitude in Fig.\,\ref{figFaddeev}, the electromagnetic current is known \cite{Oettel:1999gc}.  The key element in constructing that current is the dressed-quark-photon vertex.  It is plain from a consideration of the Ward-Green-Takahashi identities \cite{Ward:1950xp,Green:1953te,Takahashi:1957xn,Takahashi:1985yz} and the structure of the functions in Eq.\,\eqref{eqSp} that the bare vertex ($\gamma_\mu$) is not a good approximation to the dressed vertex for $Q^2 \lesssim 2\,$GeV$^2$, where (as above) $Q$ is the incoming photon momentum.  This has long been clear \cite{Ball:1980ay} and recent years have produced a sophisticated understanding of the coupling between the photon and a dressed-fermion.  Two model-independent results, which have emerged from the vast body of literature, are crucial herein \cite{Chang:2010hb,Qin:2013}: the \emph{Ansatz} described in Ref.\,\cite{Ball:1980ay} is the unique form for the solution of the longitudinal Ward-Green-Takahashi identity; and the transverse part of the dressed vertex expresses a dressed-quark anomalous magnetic moment distribution, which is large at infrared momenta.  Stated simply, the photon to dressed-quark coupling is markedly different from that of a pointlike Dirac fermion.

\begin{figure}[t]
\centerline{%
\includegraphics[clip,width=0.85\linewidth]{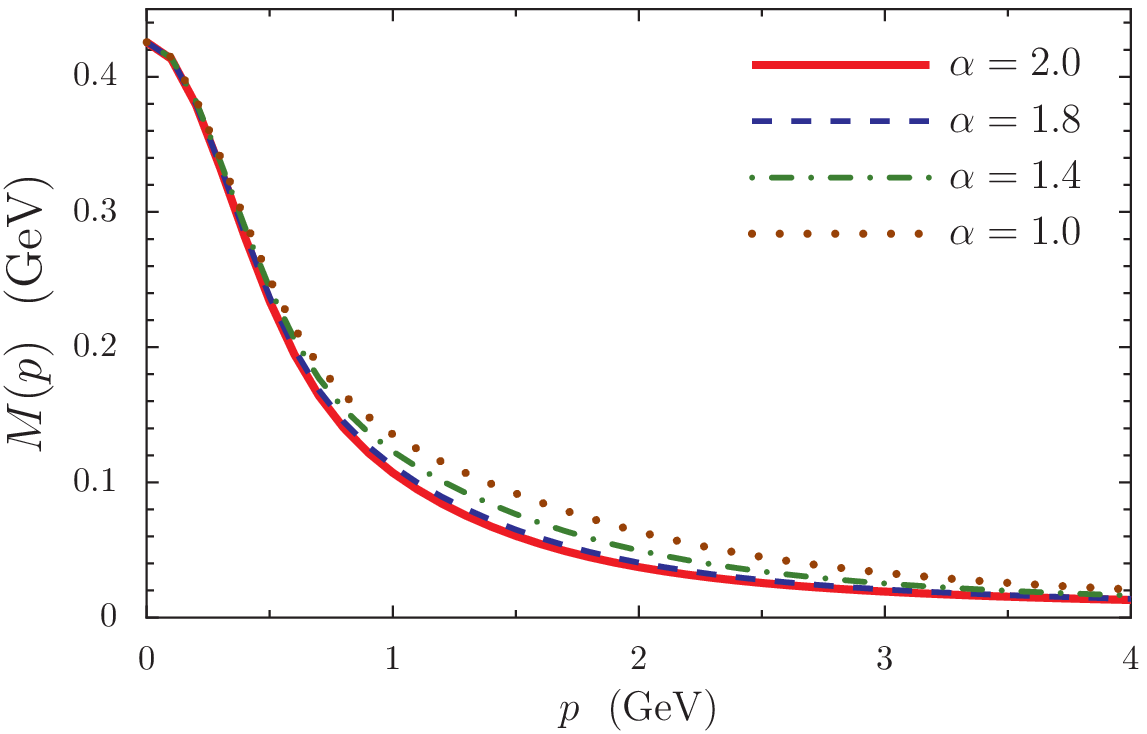}}
\centerline{%
\includegraphics[clip,width=0.85\linewidth]{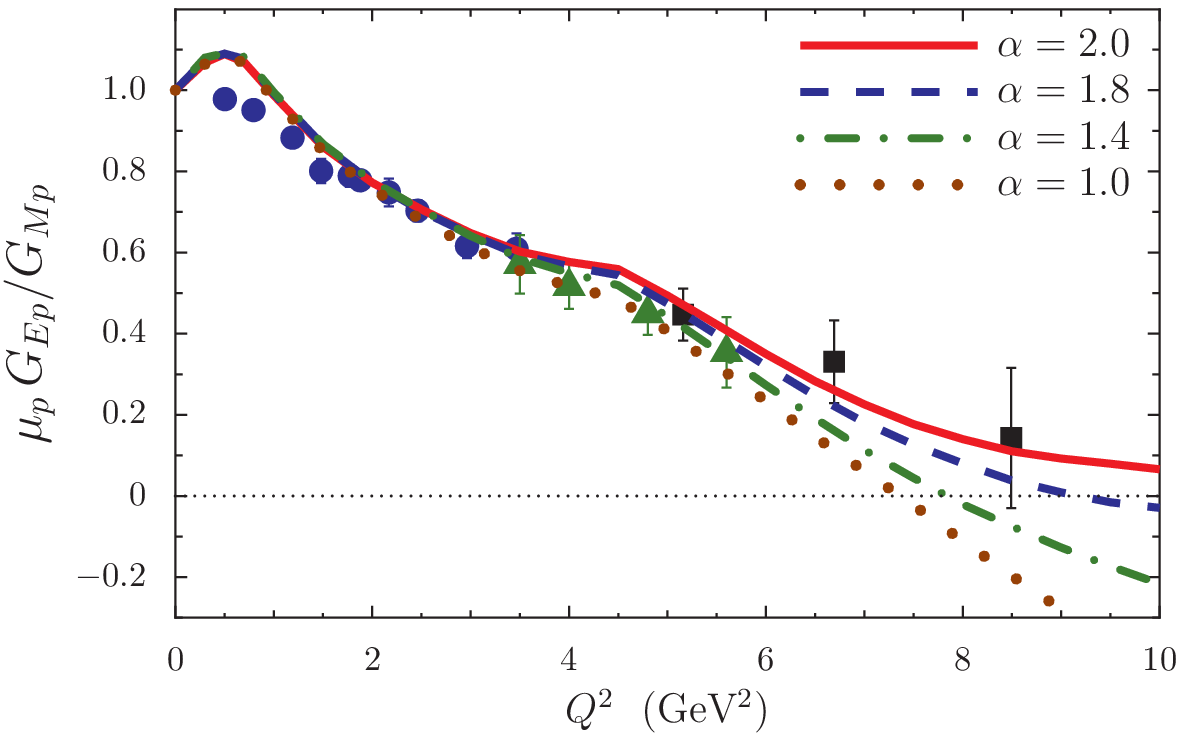}}
\caption{\label{figMp}
\emph{Upper panel}. Dressed-quark mass function.  $\alpha=1$ specifies the reference form  
and increasing $\alpha$ diminishes the domain upon which DCSB is active.
\emph{Lower panel}. Response of $\mu_p G_E/G_M$ to increasing $\alpha$; i.e., to an increasingly rapid transition between constituent- and parton-like behaviour of the dressed-quarks.  Data are from Refs.\,\protect\cite{Jones:1999rz,Gayou:2001qd,Gayou:2001qt,Puckett:2010ac,Puckett:2011xg}.}
\end{figure}

The computation of the proton's elastic form factors, using the elements detailed above, is exemplified in Refs.\,\cite{Cloet:2008re,Chang:2011tx,Cloet:2011qu}.  We use that framework herein, with the dressed-quark mass-function illustrated in the upper panel of Fig.\,\ref{figMp}, the associated dressed-quark propagator, and the following dressed-quark--photon vertex:
\begin{eqnarray}
\label{eqVertex}
\Gamma_\mu(k,p) & = & \Gamma_\mu^{\rm BC}(k,p) - \varsigma \sigma_{\mu\nu} q_\nu \Delta_B(k^2,p^2)\,,
\end{eqnarray}
with $q=k-p$, $t=k+p$, and \cite{Ball:1980ay}
\begin{eqnarray}
\Gamma_\mu^{\rm BC}(k,p) &=& \sum_{j=1}^3 \lambda_j(k,p) \, L^j_\mu(k,p)\,,
\end{eqnarray}
where: $L^1_\mu = \gamma_\mu$, $L^2_\mu= (1/2)\, t_\mu\, \gamma\cdot t$,
$L^3_\mu = - i t_\mu\, \mathbf{I}_{\rm D}$; $\lambda_1 = \Sigma_A(k^2,p^2)$, $\lambda_2 = \Delta_A(k^2,p^2)$, $\lambda_3 = \Delta_B(k^2,p^2)$; and $\Sigma_{\phi}(k^2,p^2) = [\phi(k^2)+\phi(p^2)]/2$, $\Delta_{\phi}(k^2,p^2) = [\phi(k^2)-\phi(p^2)]/[k^2-p^2]$,
with $A$, $B$ in Eq.\,\eqref{eqSp}.
The second term in Eq.\,\eqref{eqVertex} expresses the momentum-dependent dressed-quark anomalous magnetic moment distribution, with $\varsigma=0.4$ being the modulating magnitude \cite{Qin:2013}.

In order to highlight a connection between DCSB and the $Q^2$-dependence of proton form factors, one may introduce a damping factor, $\alpha$, into the dressed-quark propagator used for all calculations in Refs.\,\cite{Cloet:2008re}.  [Explicitly, we write $b_3 \to \alpha b_3$ in Eq.\,(A.19) of Ref.\,\protect\cite{Cloet:2008re}, the effect of which is a modification in Eq.\,\eqref{eqSp}
that may be approximated as $B(p)\to B(p)(1+ \alpha f(p))/(1+\alpha^2 f(p))$, $f(p)=2(p/2)^4/(1+(p/2)^6)$.]
The value $\alpha=1$ specifies the reference form of the dressed-quark propagator, which was obtained in a fit to a diverse array of pion properties \protect\cite{Burden:1995ve}.  It produces a chiral-limit condensate \cite{Brodsky:2010xf,Chang:2011mu,Brodsky:2012ku} $\langle \bar q q \rangle_0^\pi = -(0.250{\rm GeV}=:\chi^\pi_0)^3$; and is associated with a prediction of the pion's valence-quark distribution function \protect\cite{Hecht:2000xa} that was recently verified empirically \protect\cite{Aicher:2010cb}.

As $\alpha$ is increased, the rate at which the dressed-quark mass function drops towards its perturbative behaviour is accelerated so that, as evident in the upper panel of Fig.\,\ref{figMp}, the strength of DCSB is diminished and the influence of explicit chiral symmetry breaking is exposed at smaller dressed-quark momenta.  This is the qualitative impact of $\alpha$ that we exploit herein.

At each value of $\alpha$, we repeated all steps in the computation detailed in Ref.\,\cite{Cloet:2008re}.  Namely, we solved the Faddeev equation to obtain the proton's mass and amplitude, and, using that material, constructed the current and computed the proton's elastic form factors.  (The scalar and axial-vector diquark masses were held fixed as $\alpha$ was varied, in which case the nucleon mass, $m_N$, drops by $<1$\% as $\alpha$ is increased from $1.0$ to $2.0$.  Since damping was deliberately implemented so that the pointwise evolution of $M(p^2)$ to its ultraviolet asymptote is accelerated without changing $M(p^2=0)$ and because the computed values of masses are primarily determined by the infrared value of mass-functions \cite{Chen:2012qr}, this is a reasonable assumption on the input and an understandable result for $m_N$.)

The effect on $G_E(Q^2)/G_M(Q^2)$, produced by suppressing DCSB, is displayed in the lower panel of Fig.\,\ref{figMp}.  The impact is striking.  For $\alpha=1$, one recovers the result in Ref.\,\cite{Chang:2011tx}, which exhibits a zero in $G_E(Q^2)$, and hence in the ratio, at $Q^2 \approx 8\,$GeV$^2$.  However, as $\alpha$ is increased, so that the strength of DCSB is damped, the zero is pushed to larger values of $Q^2$, until it disappears completely at $\alpha=2.0$.  Associating the curves in the upper and lower panels of the figure, one observes that apparently modest changes in the rate at which the mass function drops toward its ultraviolet asymptote have a dramatic effect on the location and existence of a zero in  $G_E(Q^2)/G_M(Q^2)$.

\begin{figure}[t]
\centerline{%
\includegraphics[clip,width=0.85\linewidth]{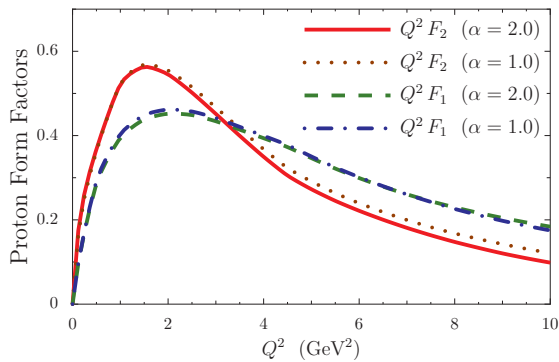}}
\caption{\label{figF2}
$Q^2$-weighted proton Dirac ($Q^2 F_1$) and Pauli ($Q^2 F_2$) form factors calculated with $\alpha=1.0$, $2.0$.  $Q^2 F_1$ shows little sensitivity to the rate at which dressed-quarks make the transition between constituent- and parton-like behaviour and hence $F_1$, none at all.  In contrast, $Q^2 F_2$, which also appears in the definition of $G_E$, exhibits a measurable dependence.}
\end{figure}

In order to explain this remarkable behaviour, it is useful to recall Eqs.\,\eqref{GEpeq}, \eqref{GMpeq}.  The magnetic form factor is a simple additive linear combination of the proton's Dirac and Pauli form factors.  Therefore, small changes in $F_{1,2}(Q^2)$, arising from the differences displayed in the upper panel of Fig.\,\ref{figMp} and illustrated in Fig.\,\ref{figF2}, cannot have a large impact.  On the other hand, the electric form factor is a \emph{difference}, in which changes in the Pauli form factor are amplified with increasing $Q^2$.

Physically, the Pauli form factor is a gauge of the distribution of magnetisation within the proton.  Absent $F_2$, the proton's electromagnetic current would be like that of a Dirac fermion.  The $Q^2=0$ value of the Pauli form factor is the proton's anomalous magnetic moment; and the evolution with $Q^2$ measures the distribution of anomalous magnetisation within the proton bound-state.  In the DSE approach, the proton's magnetisation is carried by dressed-quarks and influenced by correlations amongst them.  The latter are expressed via the Faddeev wave-function, obtained by reattaching the quark lines to the Faddeev amplitude.  This wave function exhibits $S$-, $P$- and $D$-wave quark orbital angular momentum correlations in the proton's rest frame \cite{Cloet:2007pi}.
The resulting nucleon mass is $1.18\,$GeV, a value which accommodates the material negative pion-loop corrections \cite{Cloet:2008re,Suzuki:2009nj}.



Suppose for a moment that quarks are described by a momentum-independent dressed-mass, as in Ref.\,\cite{Wilson:2011aa}.  In that counterpoint to QCD, the dressed-quarks produce hard Dirac and Pauli form factors, which yield a ratio $\mu_p G_E/G_M$ that possesses a zero at $Q^2\lesssim 4\,$GeV$^2$.

Alternatively, consider a proton comprised of dressed-quarks associated with the mass function in the upper panel of Fig.\,\ref{figMp}.  This mass function is large at infrared momenta but approaches the current-quark mass as the momentum of the dressed-quark increases.  As we have explained, such is the behaviour in QCD: dressed-quarks are massive in the infrared but become parton-like in the ultraviolet, characterised thereupon by a mass function that is modulated by the current-quark mass.  In this case, the proton's dressed-quarks possess constituent-quark-like masses at small momenta.  Thus, for all considered values of $\alpha$: these quarks possess a large anomalous magnetic moment at infrared momenta (in keeping with their large mass) \cite{Chang:2010hb}; $F_{1,2}(Q^2)$ are insensitive to $\alpha$ on this domain; and hence so is the ratio $\mu_p G_E/G_M$.

On the other hand, as the momentum transfer grows, the structure of the integrands in the computation of the elastic form factors ensures that the dressed-quark mass functions are increasingly sampled within the domain upon which the chiral condensate \cite{Brodsky:2010xf,Chang:2011mu,Brodsky:2012ku} modulates the magnitude of $M(p^2)$.  This corresponds empirically to momentum transfers $Q^2 \gtrsim 5\,$GeV$^2$.  Plainly, as this chiral order parameter becomes smaller, a part of DCSB is suppressed, and the dressed-quarks become increasingly parton-like; viz., they are partially unclothed and come to behave as light fermion degrees of freedom on a larger momentum domain.  
Following in large part, then, from the fact that light-quarks must have a small anomalous magnetic moment \cite{Chang:2010hb}, the proton Pauli form factor generated dynamically therewith drops more rapidly to zero: the quark angular momentum correlations remain but the individual dressed-quark magnetic moments diminish markedly.  This is apparent in Fig.\,\ref{figF2}.

Thus, as a consequence of suppressing the domain upon which DCSB is active, an effect expressed via a suppression of $\chi_0^\pi$ in the model used for this illustration, the zero in the ratio $\mu_p G_E/G_M$ is pushed to larger values of $Q^2$, until it disappears from the currently accessible experimental domain when $\chi_0^\pi$ falls to roughly 80\% of its unperturbed value.  Indeed, in this case there is no zero in the computed result on $Q^2>0$.

An improvement of our study is possible; e.g., via the \emph{ab initio} treatment of the DSEs detailed in Refs.\,\cite{Eichmann:2008ef,Eichmann:2011vu}.  However, close inspection of results already obtained in that more sophisticated approach lends support to our conclusions; namely, as $\chi_0^\pi/M(0)$ decreases, the proton's electric form factor approaches zero less rapidly.

We explained that the fully-consistent treatment of a quark-quark interaction which yields dressed-quarks with a constant mass-function, produces a zero in $\mu_p G_{Ep}(Q^2)/G_{Mp}(Q^2)$ at a small value of $Q^2$.  At the other extreme, a theory in which the mass-function rapidly becomes partonic -- namely, is very soft -- produces no zero at all.  From a theoretical perspective, there are numerous possibilities in between.
It follows that the possible existence and location of the zero in the ratio of proton elastic form factors [$\mu_p G_{Ep}(Q^2)/G_{Mp}(Q^2)$] are a fairly direct measure of the nature of the quark-quark interaction in the Standard Model.  They are a cumulative gauge of the momentum dependence of the interaction, the transition between the associated theory's nonperturbative and perturbative domains, and the width of that domain.  Hence, in extending experimental measurements of this ratio, and thereby the proton's charge form factor, to larger momentum transfers; i.e., in reliably determining the proton's charge distribution, there is an extraordinary opportunity for a constructive dialogue between experiment and theory.  That feedback will enable substantial progress in contemporary efforts to reveal the character of the strongly interacting part of the Standard Model and its emergent phenomena.


%
Work supported by:
University of Adelaide and Australian Research Council through grant no.~FL0992247;
%
and Department of Energy, Office of Nuclear Physics, contract no.~DE-AC02-06CH11357.


\begin{thebibliography}{10}

\bibitem{Sachs:1962zzc}
R.~Sachs,
\newblock Phys. Rev. {\bf 126}, 2256 (1962).

\bibitem{Hofstadter:1955ae}
R.~Hofstadter and R.~W. McAllister,
\newblock Phys. Rev. {\bf 98}, 217 (1955).

\bibitem{Rosenbluth:1950yq}
M.~Rosenbluth,
\newblock Phys. Rev. {\bf 79}, 615 (1950).

\bibitem{Arrington:2006zm}
J.~Arrington, C.~D. Roberts and J.~M. Zanotti,
\newblock J. Phys. {\bf G34}, S23 (2007).

\bibitem{Perdrisat:2006hj}
C.~F. Perdrisat, V.~Punjabi and M.~Vanderhaeghen,
\newblock Prog. Part. Nucl. Phys. {\bf 59}, 694 (2007).

\bibitem{Jones:1999rz}
M.~K. Jones {\em et~al.},
\newblock Phys. Rev. Lett. {\bf 84}, 1398 (2000).

\bibitem{Akhiezer:1974em}
A.~Akhiezer and M.~Rekalo,
\newblock Sov. J. Part. Nucl. {\bf 4}, 277 (1974),
\newblock [Fiz. Elem. Chast. Atom. Yadra \textbf{4}, 662 (1973)].

\bibitem{Arnold:1980zj}
R.~Arnold, C.~E. Carlson and F.~Gross,
\newblock Phys. Rev. {\bf C23}, 363 (1981).

\bibitem{Gayou:2001qd}
O.~Gayou {\em et~al.},
\newblock Phys. Rev. Lett. {\bf 88}, 092301 (2002).

\bibitem{Gayou:2001qt}
O.~Gayou {\em et~al.},
\newblock Phys. Rev. {\bf C64}, 038202 (2001).

\bibitem{Puckett:2010ac}
A.~J.~R. Puckett {\em et~al.},
\newblock Phys. Rev. Lett. {\bf 104}, 242301 (2010).

\bibitem{Puckett:2011xg}
A.~Puckett {\em et~al.},
\newblock Phys. Rev. {\bf C85}, 045203 (2012).

\bibitem{Kelly:2004hm}
J.~J. Kelly,
\newblock Phys. Rev. {\bf C70}, 068202 (2004).

\bibitem{Bradford:2006yz}
R.~Bradford, A.~Bodek, H.~S. Budd and J.~Arrington,
\newblock Nucl. Phys. Proc. Suppl. {\bf 159}, 127 (2006).

\bibitem{Arrington:2011dn}
J.~Arrington, P.~Blunden and W.~Melnitchouk,
\newblock Prog. Part. Nucl. Phys. {\bf 66}, 782 (2011).

\bibitem{Dugger:2009pn}
M.~Dugger {\em et~al.},
\newblock Phys. Rev. {\bf C79}, 065206 (2009).

\bibitem{Aznauryan:2009mx}
I.~Aznauryan {\em et~al.},
\newblock Phys. Rev. {\bf C80}, 055203 (2009).

\bibitem{Aznauryan:2011td}
I.~Aznauryan, V.~Burkert and V.~Mokeev,
\newblock AIP Conf. Proc. {\bf 1432}, 68 (2012).

\bibitem{Bashir:2012fs}
A.~Bashir {\em et~al.},
\newblock Commun. Theor. Phys. {\bf 58}, 79 (2012).

\bibitem{Bhagwat:2006tu}
M.~S. Bhagwat and P.~C. Tandy,
\newblock AIP Conf. Proc. {\bf 842}, 225 (2006).

\bibitem{Roberts:2007ji}
C.~D. Roberts,
\newblock Prog. Part. Nucl. Phys. {\bf 61}, 50 (2008).

\bibitem{Bowman:2005vx}
P.~O. Bowman {\em et~al.},
\newblock Phys. Rev. {\bf D71}, 054507 (2005).

\bibitem{national2012Nuclear}
{The Committee on the Assessment of and Outlook for Nuclear Physics; Board on
  Physics and Astronomy; Division on Engineering and Physical Sciences;
  National Research Council},
\newblock {\em Nuclear Physics: Exploring the Heart of Matter} (National
  Academies Press, 2012).

\bibitem{Cahill:1988dx}
R.~T. Cahill, C.~D. Roberts and J.~Praschifka,
\newblock Austral. J. Phys. {\bf 42}, 129 (1989).

\bibitem{Cahill:1987qr}
R.~T. Cahill, C.~D. Roberts and J.~Praschifka,
\newblock Phys. Rev. {\bf D36}, 2804 (1987).

\bibitem{Maris:2002yu}
P.~Maris,
\newblock Few Body Syst. {\bf 32}, 41 (2002).

\bibitem{Cloet:2008re}
I.~C. Clo{\"e}t, G.~Eichmann, B.~El-Bennich, T.~Kl{\"a}hn and C.~D. Roberts,
\newblock Few Body Syst. {\bf 46}, 1 (2009).

\bibitem{Chang:2011tx}
L.~Chang, I.~C. Clo{\"e}t, C.~D. Roberts and H.~L.~L. Roberts,
\newblock AIP Conf. Proc. {\bf 1354}, 110 (2011).

\bibitem{Cloet:2011qu}
I.~C. Clo{\"e}t, C.~D. Roberts and D.~J. Wilson,
\newblock AIP Conf. Proc. {\bf 1388}, 121 (2011).

\bibitem{Eichmann:2009qa}
G.~Eichmann, R.~Alkofer, A.~Krassnigg and D.~Nicmorus,
\newblock Phys. Rev. Lett. {\bf 104}, 201601 (2010).

\bibitem{Close:1988br}
F.~E. Close and A.~W. Thomas,
\newblock Phys. Lett. {\bf B212}, 227 (1988).

\bibitem{Cloet:2005pp}
I.~Clo{\"e}t, W.~Bentz and A.~W. Thomas,
\newblock Phys. Lett. {\bf B621}, 246 (2005).

\bibitem{Wilson:2011aa}
D.~J. Wilson, I.~C. Clo{\"e}t, L.~Chang and C.~D. Roberts,
\newblock Phys. Rev. {\bf C85}, 025205 (2012).

\bibitem{Cates:2011pz}
G.~Cates, C.~de~Jager, S.~Riordan and B.~Wojtsekhowski,
\newblock Phys. Rev. Lett. {\bf 106}, 252003 (2011).

\bibitem{Cloet:2012cy}
I.~C. Clo{\"e}t and G.~A. Miller,
\newblock Phys. Rev. {\bf C86}, 015208 (2012).

\bibitem{Oettel:1999gc}
M.~Oettel, M.~Pichowsky and L.~von Smekal,
\newblock Eur. Phys. J. {\bf A8}, 251 (2000).

\bibitem{Ward:1950xp}
J.~C. Ward,
\newblock Phys. Rev. {\bf 78}, 182 (1950).

\bibitem{Green:1953te}
H.~Green,
\newblock Proc. Phys. Soc. {\bf A66}, 873 (1953).

\bibitem{Takahashi:1957xn}
Y.~Takahashi,
\newblock Nuovo Cim. {\bf 6}, 371 (1957).

\bibitem{Takahashi:1985yz}
Y.~Takahashi,
\newblock (1985),
\newblock {\emph{Canonical quantization and generalized Ward relations:
  Foundation of nonperturbative approach}, Print-85-0421 (Alberta)}.

\bibitem{Ball:1980ay}
J.~S. Ball and T.-W. Chiu,
\newblock Phys. Rev. {\bf D22}, 2542 (1980).

\bibitem{Chang:2010hb}
L.~Chang, Y.-X. Liu and C.~D. Roberts,
\newblock Phys. Rev. Lett. {\bf 106}, 072001 (2011).

\bibitem{Qin:2013}
S.-x. Qin, L.~Chang, Y.-x. Liu, C.~D. Roberts and S.~M. Schmidt,
\newblock arXiv:1302.3276 [nucl-th],
\newblock \emph{Practical corollaries of transverse Ward-Green-Takahashi
  identities}.

\bibitem{Burden:1995ve}
C.~J. Burden, C.~D. Roberts and M.~J. Thomson,
\newblock Phys. Lett. {\bf B371}, 163 (1996).

\bibitem{Brodsky:2010xf}
S.~J. Brodsky, C.~D. Roberts, R.~Shrock and P.~C. Tandy,
\newblock Phys. Rev. {\bf C82}, 022201(R) (2010).

\bibitem{Chang:2011mu}
L.~Chang, C.~D. Roberts and P.~C. Tandy,
\newblock Phys. Rev. {\bf C85}, 012201(R) (2012).

\bibitem{Brodsky:2012ku}
S.~J. Brodsky, C.~D. Roberts, R.~Shrock and P.~C. Tandy,
\newblock Phys. Rev. {\bf C{\,}85}, 065202 (2012).

\bibitem{Hecht:2000xa}
M.~B. Hecht, C.~D. Roberts and S.~M. Schmidt,
\newblock Phys. Rev. {\bf C63}, 025213 (2001).

\bibitem{Aicher:2010cb}
M.~Aicher, A.~Sch{\"a}fer and W.~Vogelsang,
\newblock Phys.\ Rev.\ Lett. {\bf 105}, 252003 (2010).

\bibitem{Chen:2012qr}
C.~Chen, L.~Chang, C.~D. Roberts, S.-L. Wan and D.~J. Wilson,
\newblock Few Body Syst. {\bf 53}, 293 (2012).

\bibitem{Cloet:2007pi}
I.~C. Clo{\"e}t, A.~Krassnigg and C.~D. Roberts,
\newblock (arXiv:0710.5746 [nucl-th]),
\newblock In {\it Proceedings} {\it of} \emph{11th} \emph{International}
  \emph{Conference} \emph{on} {Meson-Nucleon Physics and} {\it the} {\it
  Structure} \emph{of} \emph{the} \emph{Nucleon} \emph{(MENU 2007)},
  J{\"u}lich, Germany, 10-14 Sep 2007, eds.\ H.~Machner and S.~Krewald, paper
  125.

\bibitem{Suzuki:2009nj}
N.~Suzuki {\em et~al.},
\newblock Phys. Rev. Lett. {\bf 104}, 042302 (2010).

\bibitem{Eichmann:2008ef}
G.~Eichmann, I.~C. Clo{\"e}t, R.~Alkofer, A.~Krassnigg and C.~D. Roberts,
\newblock Phys. Rev. {\bf C79}, 012202 (2009).

\bibitem{Eichmann:2011vu}
G.~Eichmann,
\newblock Phys. Rev. {\bf D84}, 014014 (2011).

\end{thebibliography}

\end{document}